\documentclass[preprintnumbers,10pt,nofootinbib]{revtex4}

\usepackage{amsmath,latexsym,amssymb,amsfonts}
\usepackage[dvips]{graphicx,color}
\usepackage{bm}

\begin{document}


\title{\textbf{\textit{Annihilation-to-nothing}: a quantum gravitational boundary condition for the Schwarzschild black hole}}

\author{
{\textsc{Mariam Bouhmadi-L\'{o}pez$^{a,b}$}}\footnote{{\tt mariam.bouhmadi@ehu.eus}},
{\textsc{Suddhasattwa Brahma$^{c,d}$}}\footnote{{\tt suddhasattwa.brahma@gmail.com}},
{\textsc{Che-Yu Chen$^{e,f}$}}\footnote{{\tt b97202056@gmail.com}},
{\textsc{Pisin Chen$^{e,f,g}$}}\footnote{{\tt pisinchen@phys.ntu.edu.tw}},
and
{\textsc{Dong-han Yeom$^{h,i}$}}\footnote{{\tt innocent.yeom@gmail.com}}
}

\affiliation{
$^{a}$\small{Department of Theoretical Physics, University of the Basque Country UPV/EHU, Bilbao 48080, Spain}\\
$^{b}$\small{IKERBASQUE, Basque Foundation for Science, Bilbao 48011, Spain}\\
$^{c}$\small{Asia Pacific Center for Theoretical Physics, Pohang 37673, Republic of Korea}\\
$^{d}$\small{Department of Physics, McGill University, Montr\'eal, QC H3A 2T8, Canada}\\
$^{e}$\small{Department of Physics and Center for Theoretical Sciences, National Taiwan University, Taipei 10617, Taiwan}\\
$^{f}$\small{Leung Center for Cosmology and Particle Astrophysics, National Taiwan University, Taipei 10617, Taiwan}\\
$^{g}$\small{Kavli Institute for Particle Astrophysics and Cosmology, SLAC National Accelerator Laboratory, Stanford University, Stanford, CA 94305, USA}\\
$^{h}$\small{Department of Physics Education, Pusan National University, Busan 46241, Republic of Korea}\\
$^{i}$\small{Research Center for Dielectric and Advanced Matter Physics, Pusan National University, Busan 46241, Republic of Korea}
}

\begin{abstract}
\noindent The interior of a static Schwarzschild metric can be written in terms of two functions, similar to some models of anisotropic cosmology. With a suitable choice of canonical variables, we solve the Wheeler-DeWitt equation (WDW) inside the horizon of a Schwarzschild black hole. By imposing classicality near the horizon, and requiring boundedness of the wave function, we get a rather generic solution of the WDW equation, whose steepest-descent solution, i.e., the ridge of the wave function, coincides nicely with the classical trajectory. However, there is an ambiguity in defining the arrow of time which leads to two possible interpretations -- (i) if there is only one arrow of time, one can infer that the steepest-descent of the wave function follows the classical trajectory throughout: coming from the event horizon and going all the way down to the singularity, while (ii) if there are two different arrows of time in two separate regimes, it can be inferred that the steepest-descent of the wave function comes inwards from the event horizon in one region while it moves outwards from the singularity in the other region, and there exists an annihilation process of these two parts of the wave function inside the horizon. Adopting the second interpretation could shed light on the information loss paradox: as time goes on, probabilities for histories that include black holes and singularities decay to zero and eventually only trivial geometries dominate.
\end{abstract}

\maketitle

\newpage

\tableofcontents

\section{Introduction}
The investigation of the interior of a black hole is one of the most interesting endeavours of fundamental physics. This problem is necessarily related to the resolution of singularities \cite{Hawking:1969sw}. Moreover, only if we can fully understand the nature of singularities, shall we be able to describe the entire history of a black hole from its formation to its evaporation. As a result, the information loss paradox may be solved due to some novel insight \cite{Hawking:1976ra}.

However, resolving a singularity is, of course, notoriously difficult. The basic reason is that this requires non-perturbative descriptions of quantum gravity, where there is no notion of a metric or even a background at a fundamental level, and hence perturbative quantum mechanical descriptions cannot typically be applied. For several decades, lots of approaches to quantum gravity have been developed and have tried to resolve black hole singularities.

One possible approach towards black hole singularity-resolution is to introduce a lump of strange matter which changes the geometry near the putative singularity \cite{Frolov:1988vj}. In some cases, a singularity can be partly ameliorated \cite{Yeom:2008qw}, resulting in the so-called regular black hole paradigm \cite{Hayward:2005gi}. This picture is fascinating, but still not fully satisfactory. Usual regular black holes should violate some assumptions of the singularity theorems, in which case further caveats are necessary \cite{Chen:2014jwq}. One typical problem is that such type of strange matter is distributed in a space-like direction. This implies that if this strange matter were to be modelled by some dynamical field variables, they would have to necessarily turn tachyonic, immediately making it unphysical \cite{Brahma:2019oal}. Also, in some cases where there are two horizons, the solutions can still suffer from a mass inflation singularity inherent to dynamical models \cite{Hong:2008mw}.

In order to overcome these problems, one argument is that such a lump of strange matter originates from quantum gravitational corrections. Particularly in the loop quantum gravity (LQG) community, there have been several proposals attempting to address the singularity problem, especially for semiclassical black holes\footnote{Although semiclassical, these solutions inherit some features of the nonperturbative full theory and are not equivalent to perturbative descriptions of gravity. For instance, the steepest-descent trajectory of the path integral of cosmological models in LQG is peaked on the `polymerized' trajectories as opposed to the usual classical ones \cite{Brahma:2018elv}.}. First, it was proposed that due to some quantum geometrical effects, the collapsing phase does not end up in a singularity, but it connects to a bouncing phase \cite{Ashtekar:2005cj}. If this is indeed the case, the two phases should be smoothly connected, but it has been argued that this would require non-trivial traces of quantum gravitational effects even outside the black hole horizon \cite{Haggard:2014rza}. Thereafter, it was shown that such quantum effects even exist at spatial infinity, where quantum gravity effects are generally believed to diminish \cite{Brahma:2018cgr}. In fact, with suitable modifications, it might be possible to construct quantum black hole models in which the transition from collapsing to bouncing phases only happens within the horizon \cite{Ashtekar:2018lag}, although there are still some grave criticisms of these models \cite{Bouhmadi-Lopez:2019hpp} \footnote{See, for instance \cite{Bojowald:2018xxu,Bodendorfer:2019cyv,BenAchour:2018khr, Alesci:2019pbs}, for some other models of loop quantum gravity black holes.}. 

In this work, we would like to follow a more conservative and traditional approach to the singularity problem by solving the Wheeler-DeWitt (WDW) equation \cite{DeWitt:1967yk,qgkiefer}. In the static limit and for a suitable slicing, the interior of a Schwarzschild black hole is similar to an anisotropic cosmological (Kantowski-Sachs)  model. The WDW equation is then a partial differential equation with two metric components. By applying `separation of variables', one can obtain an analytic closed form for the solution \cite{Cavaglia:1994yc}. In this paper, we propose some particular choices of boundary conditions such that the singularity can be removed. In analogy with the famous `creation-from-nothing' quantum cosmological model \cite{Hartle:1983ai}, we introduce the synonymous terminology \textit{annihilation-to-nothing} interpretation to indicate that the probability of the formation of the black hole singularity decays to zero. As we will show, if indeed such an interpretation about the wave function is correct, then this quantum gravitational treatment will shed some light on solving the information loss problem.

This paper is organized as follows. In Sec.~\ref{sec:whe}, we introduce the WDW equation for the interior of a Schwarzschild black hole and its analytic solutions. In Sec.~\ref{sec:bou}, we discuss in detail some possible boundary conditions for obtaining specific solutions. In Sec.~\ref{sec:int}, we first give an interpretation for a wave function which we will call the annihilation-to-nothing interpretation. We will show how this solution could help to shed some light on the information loss problem. Finally, in Sec.~\ref{sec:con}, we summarize and point out possible future applications of the framework we have introduced.

\section{\label{sec:whe}Wheeler-DeWitt equation inside the Schwarzschild black hole}

\subsection{Classical solution}
The interior spacetime of a static and spherically symmetric black hole can be described by a metric which is the same as that for an anisotropic cosmnological model \cite{Kantowski:1966te,Halliwell:1988wc,Laflamme:1986bc}: 
\begin{equation}
ds^{2} = - N^{2}(t) dt^{2} + a^{2}(t) dR^{2} + \frac{r_{s}^{2} b^{2}(t)}{a^{2}(t)} d\Omega_{2}^{2}\,,\label{KSmetric}
\end{equation}
where $N(t)$ is the lapse function and $-\infty < R < +\infty$. The metric components $a(t)$ and $b(t)$ are dimensionless functions of the time-like variable $t$ (in the cosmological context, these would be the so-called scale factors). The constant $r_s$ stands for the radius of the event horizon and it relates to the mass $M$ of the black hole via $r_s=2M$.

\begin{figure}
\begin{center}
\graphicspath{{fig/}}
\includegraphics[scale=0.7]{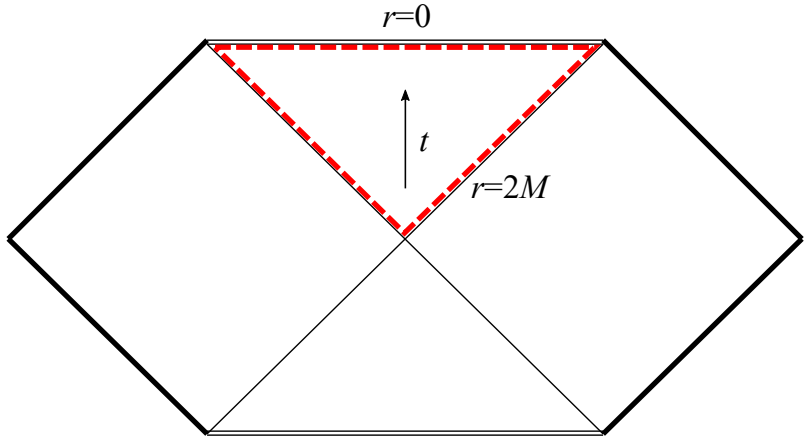}
\caption{\label{fig:pen1}The Penrose diagram of a Schwarzschild black hole. Our quantization will be applied inside the horizon (red dashed triangle). In terms of the coordinates $X$ and $Y$, the singularity; i.e., $r=0$ is located at $X=-Y\rightarrow\infty$. On the other hand, the horizon; i.e., $r=r_s=2M$ is located at $X=Y\rightarrow-\infty$.}
\end{center}
\end{figure}

For a Schwarzschild black hole, the well-known interior solution is given by
\begin{equation}
ds^{2} = - \left(\frac{r_s}{r} - 1 \right)^{-1} dr^{2} + \left(\frac{r_s}{r} - 1 \right) dR^{2} + r^{2} d\Omega_{2}^{2}\,.\label{schwarzchild2}
\end{equation}
In the metric \eqref{schwarzchild2}, the $r$ coordinate is time-like while the $R$ coordinate is space-like. Furthermore, we can redefine the time variable as follows:
\begin{equation}
r(t) = r_{s} \cos^{2} \left(\frac{t}{2}\right)\,,\qquad 0\le t\le\pi\,,
\end{equation}
which transforms the metric \eqref{schwarzchild2} to the form
\begin{equation}
ds^{2} = - r_{s}^{2} \cos^{4} \left(\frac{t}{2}\right) dt^{2} + \tan^{2} \left(\frac{t}{2}\right) dR^{2} + r_{s}^{2} \cos^{4} \left(\frac{t}{2}\right) d\Omega_{2}^{2}\,.\label{Classical}
\end{equation}
Therefore, the Schwarzschild solution given in Eq.~\eqref{schwarzchild2} and expressed in Eq.~\eqref{Classical} corresponds to the Kantowski-Sachs metric \eqref{KSmetric} where
\begin{equation}
a(t)=\tan \left(\frac{t}{2}\right)\,,\qquad b(t)=\frac{1}{2}\sin t\, \,,\qquad N(t) = r_{s} \cos^{2} \frac{t}{2}\,.\nonumber
\end{equation}
For convenience, we then define
\begin{equation}
X(t) = \ln \left(\tan \frac{t}{2}\right)\,,\qquad Y(t) = \ln \left(\frac{1}{2} \sin t\right)\,,\label{XtYt}
\end{equation}
where $X \equiv \ln a$ and $Y \equiv \ln b$. Fig.~\ref{fig:pen1} shows the Penrose diagram of a Schwarzschild black hole. One can regard this solution as a classical trajectory on the $X$-$Y$ plane, which can be expressed as follows:
\begin{eqnarray}
Y = - \ln \left( e^{X} + e^{-X} \right)\,.\label{classtraj}
\end{eqnarray}
Hence, in the $Y \rightarrow - \infty$ limit, the trajectory can be approximated either as $X = Y$ or $X = -Y$, where the former corresponds to the horizon ($t=0$), while the latter corresponds to the singularity ($t=\pi$). Although different choices of lapse functions correspond to different parametrizations of $X$ and $Y$,  it should be emphasized that the relation \eqref{classtraj} is independent of the choice of the lapse function. The functions $X(t)$ and $Y(t)$ given in Eq.~\eqref{XtYt} simply indicate one parametrization defined by the corresponding lapse function $N(t)$ from Eq.~\eqref{Classical}.

\subsection{Hamiltonian constraint and the Wheeler-DeWitt equation}
In order to derive the WDW equation describing the quantum evolution of the interior of the Schwarzschild black hole, we start with the vacuum Einstein-Hilbert action and consider the classical Hamiltonian constraint $\mathcal{H}=0$. According to the nomenclature of Dirac, the Hamiltonian constraint is a first class constraint. After promoting the classical Hamiltonian to a quantum operator, the WDW equation is derived by regarding the Hamiltonian constraint as a restriction on the Hilbert space\footnote{Here, we are speaking of the fiducial (or, kinematical) Hilbert space before solving for the WDW constraint. Formally, it is possible to write down a Hilbert space in terms of our canonically conjugate pairs although the definition of an inner product on it can be tricky (please see Sec-III C an Sec-III D for a detailed discussion).} on which the \textit{physical} wave function $|\Psi\rangle$ is defined, that is $\hat{\mathcal{H}}|\Psi\rangle=0$, where the hat denotes a quantum operator on the Hilbert space. 

The Einstein-Hilbert action is linear in the Ricci scalar $R$. Using the anisotropic metric Eq.~\eqref{KSmetric}, the Ricci scalar reads
\begin{equation}
R=\frac{2a^2}{r_s^2b^2}+\frac{2}{N^2}\left(\frac{3\dot{a}^2}{a^2}+\frac{\dot{b}^2}{b^2}-\frac{4\dot{a}\dot{b}}{ab}+\frac{\dot{a}\dot{N}}{aN}-\frac{2\dot{b}\dot{N}}{bN}-\frac{\ddot{a}}{a}+\frac{2\ddot{b}}{b}\right)\,,
\end{equation}
where the dot denotes the derivative with respect to $t$. After an integration by parts, the reduced Lagrangian of the vacuum Einstein-Hilbert action can be written as\footnote{The integration by parts can give rise to some boundary terms, which for anisotropic minisuperspace models such as these, can be canceled by adding appropriate surface terms to the Einstein-Hilbert action and has, therefore, been omitted here.}
\begin{equation}
\mathcal{L}=\frac{r_s^2b^2}{Na}\left(\frac{N^2a^2}{r_s^2b^2}+\frac{\dot{a}^2}{a^2}-\frac{\dot{b}^2}{b^2}\right)\,.
\end{equation}
From the reduced Lagrangian, one can derive the conjugate momenta
\begin{equation}
p_N=0\,,\qquad p_a=\frac{2r_s^2b^2}{Na^3}\dot{a}\,,\qquad p_b=-\frac{2r_s^2}{Na}\dot{b}\,.
\end{equation}
It can be seen that $p_N$ is a primary constraint of the system. The Hamiltonian can be constructed as follows
\begin{equation}
\mathcal{H}= \frac{N}{r_{s}^{2}}\left(\frac{a^3}{4b^2}p_a^2-\frac{a}{4}p_b^2-ar_s^2\right)+\lambda_N \frac{p_N}{r_{s}^{2}}\,,
\end{equation}
where $\lambda_N$ is a Lagrange multiplier of the constraint $p_N$. The secondary constraint associated with $p_N$ is the Hamiltonian constraint. It can be shown that $p_N=0$ and $\mathcal{H}=0$ are first class constraints and the constraint on $p_N$ essentially links to a gauge degree of freedom. The most common way to fix the gauge is by assuming a constant lapse function. In this regard, $\lambda_N$ vanishes because of the gauge fixing condition.

As has been just mentioned, the WDW equation is obtained by treating the Hamiltonian constraint as a restriction on the Hilbert space. Therefore, in the basis of $a$ and $b$, the WDW equation can be written as follows
\begin{equation}
\langle a,b|\hat{\mathcal{H}}|\Psi\rangle=\langle a,b|\frac{Na}{4b^2}\left(a^2\hat{p}_a^2-b^2\hat{p}_b^2-4r_s^2b^2\right)|\Psi\rangle=0\,.
\end{equation}
Then, we can choose the following factor orderings which is the Laplace-Beltrami ordering of derivatives with respect to the DeWitt metric on minisuperspace \cite{Laflamme:1986bc}:
\begin{equation}
a\frac{\partial}{\partial a}\equiv\frac{\partial}{\partial X}\,,\qquad b\frac{\partial}{\partial b}\equiv\frac{\partial}{\partial Y}\,,
\end{equation}
such that $X\equiv\ln a$ and $Y\equiv\ln b$. Here, because $0 \leq a, b \leq \infty$, the range of $X$ and $Y$ is $-\infty \leq X, Y \leq + \infty$. As a result, the WDW equation reads
\begin{equation}
\left(\frac{\partial^2}{\partial X^2}-\frac{\partial^2}{\partial Y^2}+4 r_s^2 e^{2Y}\right)\Psi(X,Y)=0\,,\label{WDWeq}
\end{equation}
where we have set $\hbar=1$. We can reinstate $\hbar$ by substituting $r_{s} \rightarrow r_{s}/\hbar$, which clearly shows that the classical limit is either $\hbar \rightarrow 0$ or $r_{s} \rightarrow \infty$.

\subsection{Generic solution of the Wheeler-DeWitt equation}
The WDW equation, Eq.~\eqref{WDWeq}, can be solved by using the separation of variables technique. After defining $\Psi = \phi(X) \psi(Y)$, the WDW equation can be separated into two ordinary differential equations:
\begin{eqnarray}
\frac{d^{2}\phi}{dX^{2}} + k^{2} \phi &=& 0\,,\\
\frac{d^{2}\psi}{dY^{2}} - 4 r_{s}^{2} e^{2Y} \psi + k^{2}\psi &=& 0\,.
\end{eqnarray}
In this sense, the $X$-direction can be regarded as a kind of a time-like direction (on the superspace), while the $Y$-direction is a kind of a space-like direction. Thus the equation for $\psi$ resembles a time-independent Schr\"odinger equation with a potential barrier $\sim e^{2Y}$, while the equation for $\phi$ denotes the time evolution. This interpretation justifies choosing $k$ to be real with the general solution given by
\begin{eqnarray}
\phi &=& e^{\pm ikX}\,,\\
\psi &=& C_{1} I_{ik}\left(2r_{s} e^{Y}\right) + C_{2} K_{ik}\left(2r_{s} e^{Y}\right)\,.\label{psi17}
\end{eqnarray}
Since in the classically forbidden region ($Y\rightarrow\infty$)\footnote{Note that $Y\rightarrow \infty$ lies in the classically forbidden region and the classical solution restricts the range of $Y$ such as $-\infty < Y < -\ln 2$. However, what we are considering is the full quantum gravitational wave function rather than a classical trajectory; if a part of the parameter space is quantum mechanically allowed in principle, then we need to consider it in our analysis. Hence, in order to discuss the boundedness of the wave function, we need to consider the full domain of $X$ and $Y$ (which span the Hilbert space), whether or not their values coincide with the regimes of validity of the classical solutions. In fact, the classical trajectories must arise as a consequence of the quantum dynamics.}, the modified Bessel function $I_{ik}$ diverges, we will first assume $C_1=0$ to respect the boundedness of the wave function. We will discuss the possibility of keeping the solution $I_{ik}$ later.

Before closing this section, it should be noted that the modified Bessel function $K_{ik}$ is a real function (because the argument of $K_{ik}$ is always real and positive) and it is symmetric up to $k \rightarrow -k$ change. Therefore, one can obtain
\begin{equation}
\int_{-\infty}^\infty f_1(k)e^{-ikX}K_{ik}\left(2r_{s} e^{Y}\right)dk+\int_{-\infty}^\infty f_2(k)e^{ikX}K_{ik}\left(2r_{s} e^{Y}\right)dk=\int_{-\infty}^\infty \left[f_1(k)+f_2(-k)\right]e^{-ikX}K_{ik}\left(2r_{s} e^{Y}\right)dk\,.
\end{equation}
By redefining $f(k)\equiv f_1(k)+f_2(-k)$, the general solution of the bounded wave function can be expressed with $e^{-ikX}$ and only one single weighting function $f(k)$ is required. Therefore, the most general wave function involving only $K_{ik}$ (defined in Eq.~\eqref{psi17}) reads: $\Psi=\int f(k)e^{-ikX}K_{ik}dk$.

\section{\label{sec:bou}Boundary conditions and physical interpretations}

In this section, we impose boundary conditions and provide physical interpretations of the wave function. Before we start the discussion, let us briefly find the key idea of our investigation from the analogy with quantum mechanics, because our case bears some similarities to that of the time-independent Schr\"odinger equation (since there is no time).

For example, one can consider a harmonic oscillator potential in a time-independent model. For this system, one can find and superpose several basis vectors. The wave function, resulting from such a superposition, will have a peak which is briefly located at $\langle x \rangle$. As we evaluate the expectation values, e.g., $\langle x \rangle$ or $\langle p \rangle$, we find a relation between $\langle x \rangle$ and $\langle p \rangle$, e.g., $\langle p \rangle = m d\langle x \rangle/dt$, i.e., in other words, it satisfies the classical equations of motion. This is, of course, just a part of the Erhenfest's theorem. So, one can (approximately) say that the peak of the wave function satisfies the classical equations of motion.

However, since our model is a two-dimensional problem, there is not an unique peak, but there are a series of peaks or, so to speak, there is a ridge. We called this the \textit{steepest-descent} of the wave function that satisfies the classical equations of motion. By investigating this steepest-descent within suitable boundary conditions, we will obtain a new interpretation of the wave function.

\subsection{Asymptotic behavior}
After imposing the boundedness condition by assuming $C_1=0$, the wave function can be written as
\begin{equation}
\Psi(X,Y) = \int_{-\infty}^{\infty} f(k) e^{-ikX} K_{ik}\left(2r_{s} e^{Y}\right) dk\,.
\end{equation}
As mentioned above, within the requirements of boundedness, this is the most general solution for the wave function inside the Schwarzschild black hole if we only consider $k\in \mathbb{R}$. The case of a complex-valued $k$ has been discussed below.

\begin{figure}
\begin{center}
\includegraphics[scale=0.7]{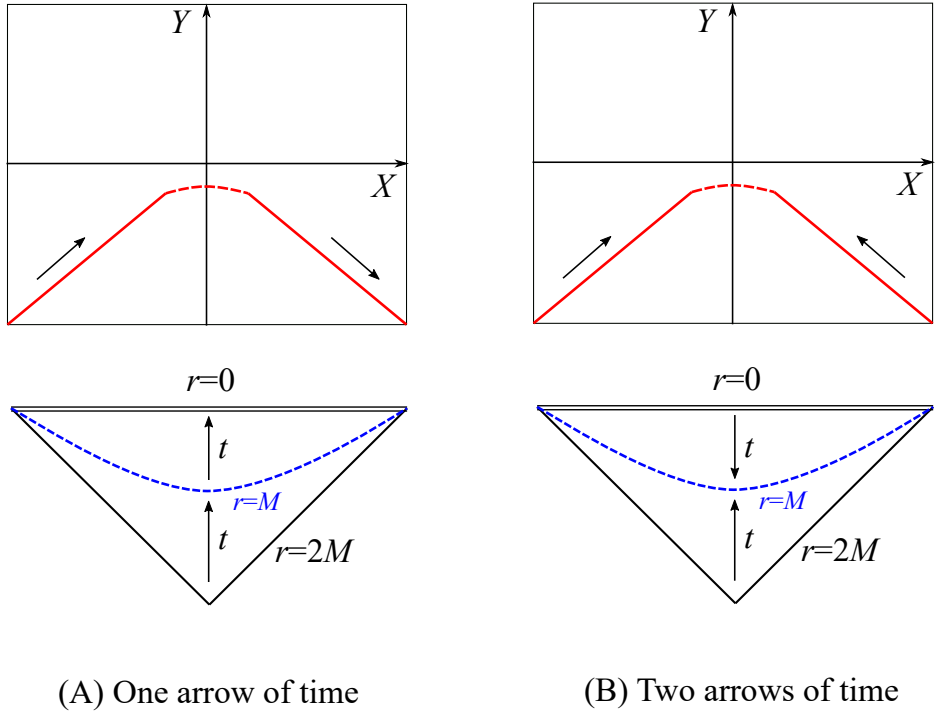}
\caption{\label{fig:pen2}There can be two interpretations: (A) there is one arrow of time and (B) there are two arrows of time in two different regimes. The upper figures denote the steepest-descents (red curve) of the wave function on the $X$-$Y$ plane. Note that the wave function vanishes at $r = M$.}
\end{center}
\end{figure}

In the $Y\rightarrow -\infty$ limit, we have
\begin{eqnarray}
K_{ik}\left(2r_{s} e^{Y}\right) \simeq \frac{1}{2} \left( r_{s}^{ik} e^{ikY} \Gamma(-ik)  + r_{s}^{-ik} e^{-ikY} \Gamma(ik)\right)\,.
\end{eqnarray}
Therefore, the most general (within our assumptions of a real-value $k$) bounded wave function in the $Y\rightarrow -\infty$ limit can be approximated as
\begin{eqnarray}
\Psi_{Y\rightarrow - \infty} = \int_{-\infty}^{\infty} \frac{f(k)}{2} \left( r_{s}^{ik} \Gamma(-ik) e^{-ik(X-Y)}  + r_{s}^{-ik} \Gamma(ik) e^{-ik(X+Y)} \right) dk\,.
\end{eqnarray}
It can be seen that this function has two peaks at $X = Y$ and $X = - Y$. Hence, the first term corresponds to near-horizon behavior, while the second term corresponds to near-singularity behavior. If there is only one arrow of time and we assume that the peak is coming inward from the event horizon (the left panel of Fig.~\ref{fig:pen2}), the incoming pulse near the horizon approaches the potential barrier near $Y = -\ln2$ and bounces to the singularity. With only one function $f(k)$ present in the wave function, there is no freedom to choose any other boundary condition at the singularity. Hence, \textit{it is impossible to choose the DeWitt boundary condition} once the boundedness condition of the wave function on the classically forbidden region is imposed. Note that here we have assumed a real $k$ in order to have a continuous spectrum and the solution acquires a wave-like behavior. If we allow $k$ to be imaginary, one can easily obtain solutions satisfying the DeWitt boundary condition at the singularity by choosing the exponentially decaying solution. However, this is beyond the scope of this paper and we leave this interesting possibility as a future research topic.

\subsection{Gaussian wave packet solution}\label{sec.3b}
\begin{figure}
\begin{center}
\includegraphics[scale=1.2]{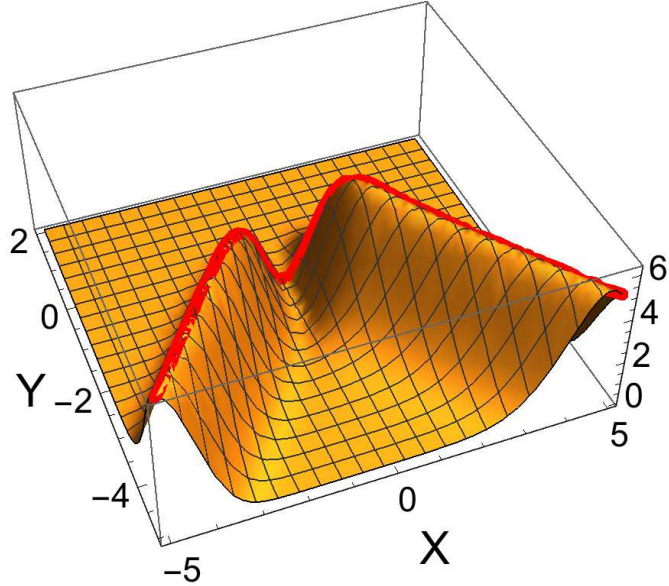}
\caption{\label{fig:wf1}The modulus squared of the wave function \eqref{wdwsoltuionnor1} is shown numerically (assuming $A=\sigma=r_{s}=1$ for simplicity). The wave number $k$ is integrated within the range $[-8,8]$. The red curve stands for the classical trajectory on the $X$-$Y$ plane.}
\end{center}
\end{figure}

In order to illustrate the behavior of the wave function more explicitly, let us choose $f(k)$ as a Gaussian wave packet such that
\begin{equation}
f(k) = \frac{2A e^{-\sigma^{2}k^{2}/2}}{\Gamma(-ik) r_{s}^{ik}}\,,
\end{equation}
where $A$ is the normalization constant and $\sigma$ is the standard deviation of the pulse at $X = Y$. Then the solution in the $X$-$Y$ plane can be written as
\begin{eqnarray}
\Psi(X,Y) = \int_{-\infty}^{\infty} \frac{2A e^{-\sigma^{2}k^{2}/2}}{\Gamma(-ik) r_{s}^{ik}} e^{-ikX} K_{ik}\left(2r_{s} e^{Y}\right) dk\,.\label{wdwsoltuionnor1}
\end{eqnarray}
The wave function, Eq.~\eqref{wdwsoltuionnor1}, on the $X$-$Y$ plane is illustrated in Fig.~\ref{fig:wf1}. The classical trajectory, Eq.~\eqref{classtraj}, is depicted by the red curve. It can be seen that there is a Gaussian wave packet at the horizon ($X = Y$ and $Y \rightarrow -\infty$), which bounces due to the potential barrier $\sim e^{2Y}$, connecting to another Gaussian wave packet near the singularity ($X = -Y$ and $Y \rightarrow -\infty$). Interestingly, there emerges a quantum bounce near $X\sim 0$ where the two Gaussian wave packets join. 

One can interpret this result as follows. The wave function has the steepest-descent, i.e., the ridge of the wave function, with an almost constant probability which coincides well with the classical trajectory (red curves of Fig.~\ref{fig:pen2} and Fig.~\ref{fig:wf1}) except at the bouncing point $X \sim 0$. Since the probability does not vary along the steepest-descent (except near $X \sim 0$), one can interpret that the steepest-descent represents a classical solution. At the level of the wave function, there is no way to define the arrow of time as can be done in the classical regime. Therefore, just as in the paradigm of quantum cosmology in the context of the no-boundary wave function \cite{Hartle:2007gi}, there is an ambiguity in choosing if there is only one arrow of time, or two.  Note that there is only one wave function and there can be the same arrow of time in all regimes or different ones in different regimes of the wave function. For the former interpretation, one can interpret that the steepest-descent trajectory is very closely followed by the wave packet as it comes inwards from the horizon, which coincides with the classical trajectory, and goes all the way down to the singularity. This is nothing but a classical black hole solution. We shall elaborate in the next section the consequences of the other possible interpretation: having two arrows of time in two different regimes.

\subsection{An analytic example: gauge choices and slicing dependence}

At this point, it would be good to comment more on the subtleties in defining the normalizability of the wave function. In order to demonstrate this, let us introduce a simple model which is analytically integrable.

In order to define the normalization, we need to choose the slicing of the hypersurface or the gauge. The problem is that there is no canonical way to choose this. It is important to emphasize that incorrect gauge choices can also lead to apparently non-normalizable wave functions \textit{even if we choose $C_1=0$} \cite{Obregon:1998jv}. For our WDW equation
\begin{eqnarray}
\left( \frac{\partial^{2}}{\partial X^{2}} - \frac{\partial^{2}}{\partial Y^{2}} + 4 r_{s}^{2} e^{2Y} \right) \Psi(X,Y) = 0\,,
\end{eqnarray}
the eigen-solutions are given by
\begin{eqnarray}
\Psi \sim e^{\pm i k X}\, K_{i k}\left(2 r_s e^{Y} \right)\,.
\end{eqnarray}
If we replace the sum over the oscillating parts of the wave function in favor of a trigonometric function, we can rewrite the wave function as \cite{Obregon:1998jv}
\begin{eqnarray}
\Psi = \int_{0}^{\infty}d k\, k \sin(k X)\, K_{ik}\left(2 r_s e^{Y}\right) \,.\label{wavefunctionexpl2}
\end{eqnarray}
Note that we have included a linear weighting function $k$ in the integral. Thus far the difference of this analysis from that in Sec.~\ref{sec.3b} is that we have not chosen an arbitrary function of $k$ in the integral to evaluate a sharply-peaked ``Gaussian'' state. Using the identity \cite{integralbook}
\begin{equation}
\int_0^\infty x\sin\left(ax\right)K_{ix}(b)dx=\frac{\pi b}{2}\sinh(a)\textrm{exp}\left[-b\cosh(a)\right]\,,\qquad |\textrm{Im}\,a|<\frac{\pi}{2}\,,\quad b>0\,,
\end{equation}
Eq.~\eqref{wavefunctionexpl2} can be calculated exactly as follows
\begin{eqnarray}
\Psi\left(X,Y\right) = \pi r_se^Y \sinh(X)\, e^{-2 r_s e^{Y}\cosh(X)}\,.
\end{eqnarray}
Note that this wave function is $\Psi \propto \tanh X$ along the steepest-descent, and its behavior is qualitatively the same as the solution \eqref{wdwsoltuionnor1} given in the previous subsection.

If we use the gauge condition that the angular part of the metric must recover the usual Schwarzschild form, i.e.,
\begin{eqnarray}
r_s^2 e^{2Y} e^{-2X} = \zeta^2\,,
\end{eqnarray}
then we can replace $e^Y = e^X\,\left(\zeta/r_s\right)$ in the wave function to get
\begin{eqnarray}
\Psi\left(X,\zeta\right) = \mathcal{N}  \left(e^{2X} - 1\right)\, \zeta e^{-\zeta \left(e^{2X} + 1\right)}\,,
\end{eqnarray}
where $\mathcal{N}$ is a numerical constant. But clearly, this solution is not normalizable under the usual (Lebesgue) measure since $\Psi$ does not go to zero when $X\rightarrow -\infty$. (On the other hand, in the limit $X\rightarrow\infty$, the wave function decays rapidly.) Although there are certain subtleties in choosing the measure on the minisuperspace Hilbert space \cite{Halliwell:1988wc}, the problem here has more to do with the fact that the wave function does not decay sufficiently fast in a particular direction rather than to do with our choice of the measure.

This is not so surprising, because the wave function is approximately $\Psi \sim \Psi(X-Y)$ for $X < 0$, while $\Psi \sim \Psi(X+Y)$ for $X > 0$; hence, the better choice of gauge for $X < 0$ is a function of $X + Y$, rather than $X - Y$. This undesirable result appears as a manifestation of the above ``gauge-fixing" condition and should not appear in general.

\subsection{Unbounded wave function}
At the end of Sec.~\ref{sec:whe}, we had disregarded the modified Bessel function $I_{ik}$ in the solution in order to maintain the boundedness of the wave function ($I_{ik}$ diverges in the classically forbidden region for $k\in\mathbb{R})$). We have shown that there is always a steepest-descent trajectory which approaches the singularity in the case where $I_{ik}$ is omitted. However, what happens if we \textit{do} allow the wave function to be unbounded? In this case, the function $I_{ik}$ can be introduced. In the $Y\rightarrow -\infty$ limit,
\begin{eqnarray}
I_{ik}\left(2r_{s} e^{Y}\right) \simeq \frac{r_{s}^{ik} e^{ikY}}{\Gamma(1+ik)}\,.
\end{eqnarray}
Naively, the generic solution can be written as
\begin{eqnarray}
\Psi(X,Y) = \int_{-\infty}^{\infty} \left [ f(k) e^{-ikX} K_{ik}\left(2r_{s} e^{Y}\right) + g(k) e^{-ikX} I_{ik}\left(2r_{s} e^{Y}\right) + h(k) e^{+ikX} I_{ik}\left(2r_{s} e^{Y}\right) \right] dk\,,\label{non22}
\end{eqnarray}
where $f(k)$, $g(k)$, and $h(k)$ are arbitrary functions. Based on the fact that the imaginary part of $I_{ik}$ is linearly dependent on $K_{ik}$, the wave function \eqref{non22} can be recast into the following concise expression:
\begin{eqnarray}
\Psi(X,Y) = \int_{-\infty}^{\infty} \left [ F(k) e^{-ikX} K_{ik}\left(2r_{s} e^{Y}\right) + G(k) e^{-ikX} M_{ik}\left(2r_{s} e^{Y}\right)  \right] dk\,,\label{non222}
\end{eqnarray}
where
\begin{eqnarray}
F(k)&\equiv& f(k)-ig(k)\frac{\sinh{\pi k}}{\pi}+ih(-k)\frac{\sinh{\pi k}}{\pi}\,,\quad \nonumber\\
G(k)&\equiv&\left[g(k)+h(-k)\right]\frac{\cosh{\pi k}}{\pi}\,,\nonumber\\
M_{ik}&\equiv&\frac{\pi}{2\cosh{\pi k}}\left(I_{ik}+I_{-ik}\right)\,.
\end{eqnarray}

\begin{figure}
\begin{center}
\includegraphics[scale=0.9]{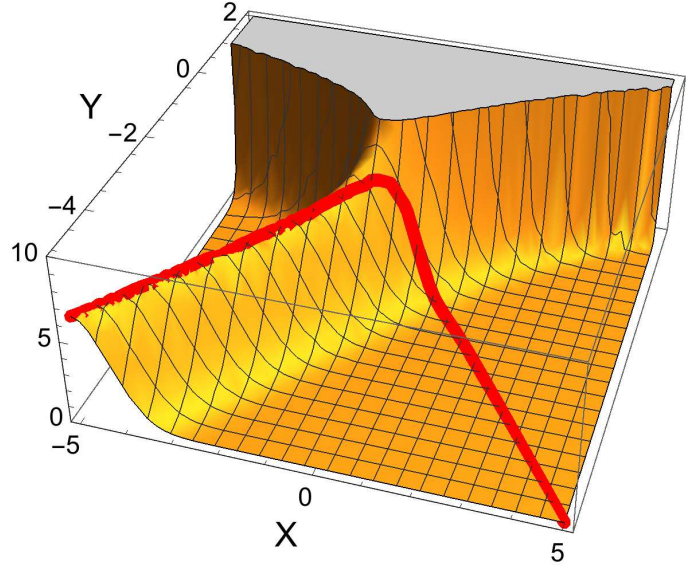}
\caption{\label{fig:wfnon2}The modulus squared of the wave function defined by Eqs.~\eqref{non222}, \eqref{non23}, and \eqref{non25} is shown numerically. The wave number $k$ is integrated within the range $[-8,8]$. The red curve stands for the classical trajectory on the $X$-$Y$ plane.}
\end{center}
\end{figure}

If we choose $g(k) = 0$ and impose the following condition
\begin{equation}
h(k) = - \frac{1}{2} \Gamma(-ik) \Gamma(1+ik) f(-k)\,,\label{non23}
\end{equation}
the wave function near $Y \rightarrow - \infty$ can be approximated as
\begin{equation}
\Psi_{Y\rightarrow - \infty} = \int_{-\infty}^{\infty} \frac{f(k)}{2} r_{s}^{ik} \Gamma(-ik) e^{-ik(X-Y)} dk\,,
\end{equation}
and the wave packet only appears near the event horizon ($X=Y\rightarrow-\infty$). For example, if we insert the Gaussian wave packet:
\begin{equation}
h(k) = - A \frac{\Gamma(-ik)}{\Gamma(ik)} \Gamma(1+ik) e^{-\sigma^{2}k^{2}/2} r_{s}^{ik}\,,\label{non25}
\end{equation}
the wave function squared on the $X$-$Y$ plane is shown in Fig.~\ref{fig:wfnon2}. It can be seen that the Gaussian wave packet only appears at the horizon, while there is no bounced wave packet toward the singularity. This is a fulfillment of the DeWitt boundary condition, although the wave function is unbounded when $Y\rightarrow\infty$. Note that an identical wave function can be constructed by choosing
\begin{equation}
f(k) = h(k) = 0\,,\qquad g(k)=A \frac{\Gamma(1+ik)}{r_s^{ik}} e^{-\sigma^{2}k^{2}/2}\,.\label{nonbound2}
\end{equation}
It should be noticed that one can also construct a wave function which vanishes near the horizon but contains a Gaussian wave packet near the singularity. This can be done by exchanging the $g(k)$ and $h(k)$ in Eq.~\eqref{nonbound2}. Naturally, this case is not of any physical interest to us.

Before ending this subsection, we should emphasize that the unboundedness of wave functions does not necessarily imply non-normalizability. This is why we have been careful not to invoke normalizability as a selection criterion for the wave function, but rather just the fact that it be bounded. In fact, due to the lack of a well-defined Hilbert space, the explicit definition of probability is still not clear in this model. When defining the inner product, there may be some measures of integration which give rise to a well-defined probability, even though the wave function diverges. However, such a measure may lack physical justification. Therefore, in the following sections, we will not investigate more about these unbounded wave functions.

\section{\label{sec:int}Interpretation}

\subsection{What if there are two arrows of time? \textit{Annihilation-to-nothing}}

\begin{figure}
\begin{center}
\includegraphics[scale=1]{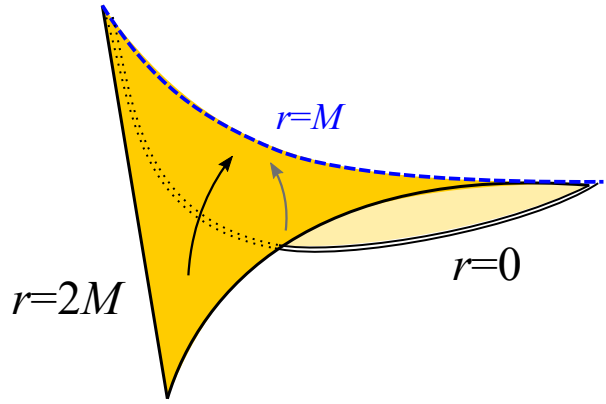}
\caption{\label{fig:pen3}If we accept the second interpretation with two arrows of time, then the spacetime associated with the Gaussian wave function coming from the horizon as well as from the singularity are both annihilated at $r \simeq M$ hypersurface.}
\end{center}
\end{figure}

Let us now turn our attention back to the bounded wave function, which is given by Eq.~\eqref{wdwsoltuionnor1} and illustrated in Fig.~\ref{fig:wf1}. The trivial interpretation is that the steepest-descent trajectory follows the classical solution of the interior of the Schwarzschild black hole (left panel of Fig.~\ref{fig:pen2}). However, due to the ambiguity of defining the arrow of time, one can also interpret that there are two arrows of time pointing towards the $r \sim M$ hypersurface, on which the quantum bounce takes place, one from the event horizon and the other from the singularity (right panel of Fig.~\ref{fig:pen2}). If we interpret the solution in this way, then as shown in Fig.~\ref{fig:wf1}, the two wave packets annihilate at the $r \sim M$ hypersurface. This is nothing but an annihilation of the spacetime geometry to nothing (Fig.~\ref{fig:pen3}), and eventually, the probability associated with the entire hypersurface will decrease to zero. Note that this annihilation-to-nothing process refers to two arrows of time for the \textit{same} wave function in different regimes.

One can provide a nice analogy of our model with a similar duality of quantum mechanics. In quantum mechanics, a particle can move forwards or backwards in time. Hence, with the same steepest-descent, the interpretations of having one arrow or two arrows of time are both possible. The exact same phenomenon also happens in quantum cosmology \cite{Hartle:2007gi}. In loop quantum cosmology, usually the big-bang and other cosmological singularities are resolved due to quantum geometrical corrections and the solution can be further extended towards the past \cite{Ashtekar:2003hd}. This is typically interpreted as a big bounce\footnote{See, however, \cite{Bojowald:2019ujl} for how the bounce, or the absence of it, depends on the quantum state.} and there is only one arrow of time. However, since the notion of time is not well-defined in the quantum gravitational realm, there is a possibility of defining two arrows of time, where this is indeed possible \cite{Brahma:2018elv}. Moreover, there are multiple indications that non-Riemannian geometry arises in loop quantum gravity due to the same effects responsible for singularity resolution where such a simple picture for a bounce has to be replaced with a more non-geometric interpretation. One of the main effects of such non-Riemannian geometry is the phenomenon of non-singular signature-change \cite{Bojowald:2016itl}, whereby one loses the usual ``time'' coordinate in the deep quantum regime, forcing one to revisit the interpretation of time in such scenarios \cite{Bojowald:2014zla}. However, the usual probabilistic picture as relevant for the no-boundary wave function, is still applicable \cite{Bojowald:2018gdt}.

Inside a black hole, on the other hand, the singularity gets resolved due to similar quantum geometrical corrections and the spacetime may be extended as a result \cite{Bojowald:2018xxu}. Once again, it does not necessarily imply the existence of only one arrow of time. In fact, it is reasonable to assume that the two arrows of time paradigm is also a viable way of interpreting even in these models. The annihilation-to-nothing interpretation is an explicit realization of this \cite{Chen:2016ask}; of course, without any loop quantum gravity effects.

\begin{figure}
\begin{center}
\includegraphics[scale=0.7]{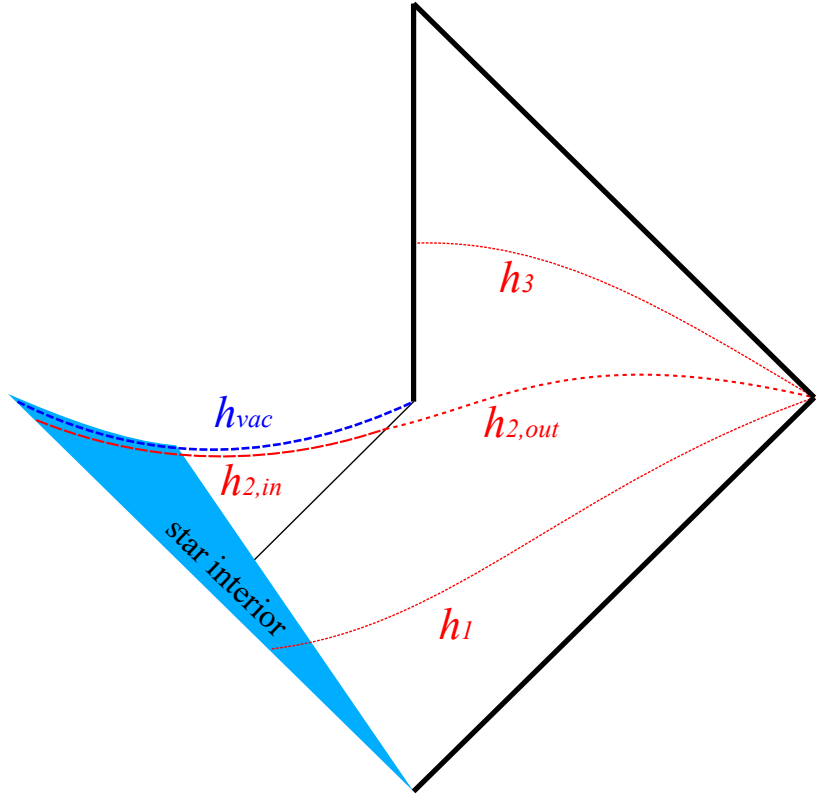}
\caption{\label{fig:pen4}Generalized causal structure of a collapsing and evaporating black hole.}
\end{center}
\end{figure}

As a speculative generalization, it is somewhat natural to expect that the same thing will happen for generic collapsing black holes (Fig.~\ref{fig:pen4}). Then, as time goes on, the space-like hypersurface will be divided into two parts; inside and outside the horizon. In the far past, the initial condition was entirely imposed on the hypersurface $h_{1}$. As time goes on, the space-like hypersurface $h_{2}$ is divided into two parts $h_{2,in}$ and $h_{2,out}$. The hypersurface will evolve as follows: $h_{1} \rightarrow h_{2,in} \cup h_{2,out} \rightarrow h_{vac} \cup h_{3}$. The probability of $h_{2,in}$ will decrease and approach zero at $h_{vac}$, where the annihilation takes place. Hence, $\Psi[h_{vac} \cup h_{3}]$ will be zero. However, if we only consider the exterior of the black hole, we have $|\Psi[h_{2,out}]| \simeq |\Psi[h_{3}]|$ and hence the observer outside the horizon will still experience a semi-classical geometry.

\begin{figure}
\begin{center}
\includegraphics[scale=0.35]{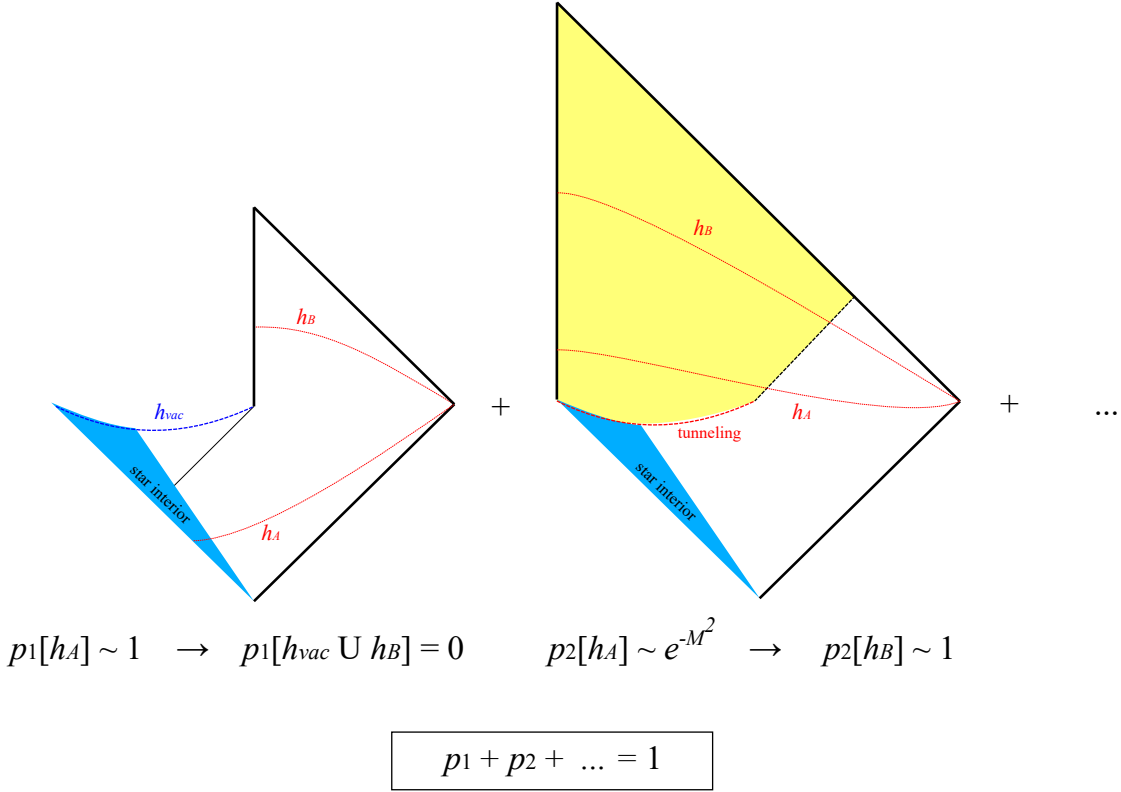}
\caption{\label{fig:pen5}Conceptual picture of the entire wave function. Let us assume that there are several dominant histories, where one (say, $p_{1}$ as its probability) is the usual semi-classical black hole and the other (say, $p_{2}$ as its probability) is a trivial geometry which includes a tunneling process. At the hypersurface $h_{A}$, the probability is dominated by the history $1$. However, as time goes on, at the hypersurface $h_{B}$, the probability is dominated by the history $2$ (if sum of all histories is one). In the end, in the entire wave function, information will be recovered by history $2$.}
\end{center}
\end{figure}

\subsection{Applications to the information loss problem}

At a first glimpse, one may suspect that the causal structure of a single universe loses information and unitarity because $\Psi[h_{vac} \cup h_{3}]$ decays to zero eventually. However, if we consider the path integral scheme and regard all the possible histories of different universes as the entire wave function, the picture gets nicely resolved.

In the entire path integral, there exists a history such that tunneling happens and the black hole disappears before a singularity is formed \cite{Sasaki:2014spa}. This tunneling is generically possible, but the tunneling probability is exponentially suppressed (Fig.~\ref{fig:pen5}). However, as we have shown, the probability of a black hole spacetime which contains a singularity will eventually decrease to zero due to the annihilation process. Therefore, in the long run, the spacetime producing a trivial geometry will dominate probabilistically subsequent to the tunneling process . In this trivial geometry, there is no loss of information \cite{Maldacena:2001kr}. Generically, information will be preserved by the entire wave function, while a semi-classical observer will experience a loss of information.

Indeed, this picture was first proposed by Maldacena and Hawking \cite{Maldacena:2001kr}, but they neither provide (i) a detailed mechanism explaining the tunneling to the trivial topology nor (ii) an explanation as to whether the contribution of the exponentially suppressed geometry is enough to preserve unitarity, e.g., the Poincare recurrence theorem may not be satisfied \cite{Bocchieri:1957}. For the first problem, we can provide a very generic mechanism for such a tunneling process \cite{Chen:2018aij}. For the second problem, this paper gives a very constructive interpretation. Due to the annihilation-to-nothing inside the horizon, eventually the probability of a black hole spacetime decreases to zero. Consequently, the contribution of the trivial geometry must dominate at late times.

This interpretation is also consistent with other observations. Firstly, the asymptotic observer will see a semi-classical black hole at the cost of losing unitarity in his view. Secondly, the superspace observer (the observer who can determine the probability of all the histories) will recover all information, but the effective geometry cannot be semi-classical. From the beginning, the superspace observer may experience a semi-classical black hole, but later, the probability will be dominated by the trivial geometry. Therefore, general relativity ceases to be valid for the superspace observer due to the superposition of various geometries \cite{Hartle:2015bna}. Consequently, there is no observer who can retain both semi-classical gravity and unitarity. If these two conditions are satisfied simultaneously, then there must be inconsistencies \cite{Yeom:2009zp}. Such a superposition of geometries can result in a violation of the classical equations of motion, which can be named as a naked firewall-like phenomenon \cite{Kim:2013fv}, although it cannot be observed by a usual semi-classical observer.

\section{\label{sec:con}Conclusion}

In this paper, we have investigated the quantum gravitational wave function for the interior of a Schwarzschild black hole. By choosing suitable canonical variables, the WDW equation can be solved analytically by the method of separation of variables.

Several possible boundary conditions for the wave function are studied in order to decide the exact form of the wave function. If we allow for an unbounded wave function, the DeWitt boundary condition for singularity avoidance can be satisfied.  Of course, in order to make such a wave function physically viable, we also need to postulate a suitable measure on the Hilbert space which is beyond the scope of this paper\footnote{Indeed, a measure is certainly required to rigorously define any wave function; however, our comment should be understood in the sense of the additional difficulty one faces in defining it for an unbounded wave function.}. On the other hand, if we require the boundedness of the wave function, we can adopt the annihilation-to-nothing interpretation, i.e., there are two arrows of time and two pieces of spacetime are annihilated at the quantum bouncing point. We emphasize that we still have only one wave function but associate two different arrows of time with it in two separate regimes. We are allowed to do so since there is no fundamental time parameter in the canonical formulation of quantum gravity. For our annihilation-to-nothing wave function, we have shown how there is a more concrete handle over the information loss problem with a possibility of resolving it. Moreover, we have also emphasized how gauge choices can become crucial in solving the WDW equation for physically relevant solutions. Of course, this is nothing new but the age-old problem of quantum ambiguities of the WDW equation reappearing in a different guise.

Our result opens up new possibilities which shall be explored in the future. Firstly, this method can be applied to other types of black holes, e.g., charged and rotating black holes. Secondly, we can check whether our analysis still holds by choosing another set of canonical variables. Finally, although in this paper we focus on the WDW quantization of the Schwarzschild black hole, this new interpretation can also be applied to other settings such as loop quantum gravity or stringy black holes. We have qualitatively extended our discussions including not only inside but also outside a black hole, but this must need further mathematical justification. In addition to them, it is fair to ask for several fundamental questions, e.g., the correct scalar product in a Hilbert space in the context of the Wheeler-DeWitt equation. This is beyond the scope of this paper, but very essential for the consistent and complete understanding of the wave function of the black hole \cite{Barvinsky:1993jf}. We are hopeful that the annihilation-to-nothing interpretation might turn out to be a viable starting point for resolving the information loss paradox.

\newpage

\section*{Acknowledgment}

The authors would like to thank anonymous referees for critical and important comments about this paper. MBL is supported by the Basque Foundation of Science Ikerbasque. She also would like to acknowledge the partial support from the Basque government Grant No. IT956-16 (Spain) and from the project FIS2017-85076-P (MINECO/AEI/FEDER, UE). CYC and PC are supported by Ministry of Science and Technology (MOST), Taiwan, through No. 107-2119-M-002-005, Leung Center for Cosmology and Particle Astrophysics (LeCosPA) of National Taiwan University, and Taiwan National Center for Theoretical Sciences (NCTS). CYC is also supported by MOST, Taiwan through No. 108-2811-M-002-682. PC is in addition supported by US Department of Energy under Contract No. DE-AC03-76SF00515. The research of SB and DY is supported in part by the Ministry of Science, ICT \& Future Planning, Gyeongsangbuk-do and Pohang City and the National Research Foundation of Korea grant no. 2018R1D1A1B07049126. SB is also supported in part by funds from NSERC, from the Canada Research Chair program and by a McGill Space Institute fellowship.

\end{document}